\documentclass[a4paper,final,12pt,onecolumn]{article}
\usepackage{setspace}
\usepackage[cp1250]{inputenc}
\usepackage{graphicx}
\usepackage{mathrsfs}
\usepackage{url}
\usepackage{amssymb}
\usepackage{amsmath}
\usepackage{amsthm,amscd,amsfonts}

\newtheorem{theorem}{Theorem}

\newcommand{\vecx}{{\bf x}}
\newcommand{\vecs}{{\bf s}}
\newcommand{\MM}{{\cal MM}}
\newcommand{\supp}{\textrm{supp}}

\title{Phoneme discrimination using $KS$-algebra I.}

\begin{document}

\author{Ondrej~\v Such\thanks{O.\v Such is with Slovak Academy of Sciences, Bansk\'a Bystrica, Slovakia, {\tt ondrejs@savbb.sk}}%
\thanks{Work on this paper was partially supported by research grant VEGA 2/0112/11}%
}%
\maketitle

\begin{abstract}
In our work we define a new algebra of operators as a substitute for fuzzy logic. Its primary purpose is for construction of binary discriminators for phonemes based on spectral content. It is optimized for design of non-parametric computational circuits, and makes uses of 4 operations: $\min$, $\max$, the difference and generalized additively homogenuous means.
\end{abstract}

\section{Introduction}

Probability, statistics and in particular Bayesian statistics are disciplined foundation for understanding uncertainty phenomena.
Fuzzy logic is a generalization of Boolean logic that purports to provide another way to describe and reason with uncertainties \cite{zad95}, \cite{zad08}. A very useful feature of fuzzy logic is that it can be implemented in analog circuits. It serves as a basis for fuzzy set theory and since its inception has found thousands of applications ranging from modelling and prediction to industrial controllers. There is really not a single fuzzy logic, but rather many variations. Most of them however use $\min$ and $\max$ operators \cite{ZavalaNieto}. Recently, a new way to compute voltage-mode $\min$ and $\max$ operations using memristors \cite{chua71} has been proposed in \cite{arxiv}. Unlike previous implementations using bipolar junction transistors \cite{Yamakawa93}, \cite{Yamakawa1}, and CMOS transistors \cite{Fatt94}, \cite{Baturone97}, the proposed circuit is passive. This has prompted research inquiries about computational power of circuits whose elements are mostly $\min$ and $\max$ operators, extending classical research on sorting networks  \cite{bat68} that use $\min$ and $\max$ exclusively.

First steps were taken in works
 \cite{Foltan}, \cite{FoltanSmiesko}, \cite{FoltanPhd}, \cite{Badura1}, \cite{BaduraPhd} where authors find pattern recognition circuits using fuzzy logic operators. Their application domain is speech recognition. In this note we lay out our arguments why we believe that  fuzzy logic may not be the best choice to use in this context. We propose an alternative that we call $KS$-algebra $\MM$. The key principles guiding our proposal are:
 \begin{itemize}
 \item simplicity, keeping the family of operations small to allow for more efficient optimization using evolutionary algorithms, which are commonly used to search for optima in discrete spaces, \cite{Sekanina1},\cite{EvolutionTexas}
 \item richness, that would allow expression of desirable properties, such as being able to capture the concept of formant \cite[page 12]{Mariani1}, or provide invariance under volume adjustment,
 \item nonlinearity, by which we mean using nonlinear operators, such as $\min$ and $\max$, as well as the absence of continuous parameters, which are currently hard to implement in analog electrical circuits.
 \end{itemize}
 
This is the first paper in a series in which we will investigate phoneme discrimination circuits using this algebra. Let us note that it is possible to use the algebra also to design circuits for feature detection (e.g. lines, edges) in vision applications, but speed of computation when simulated on digital computers is not competitive.

\section{Basic properties of human speech processing}

In order to design an efficient speech recognition framework it is helpful to understand underlying mechanisms that allow humans process audio signal. If the framework takes the mechanisms into acccount, it is more likely to exhibit performance closer to humans. This is demonstrated by improved performance of mel-based cepstral coefficients, PLP \cite{Hermansky90} or RASTA \cite{Hermansky92}.

From physical point of view sound is simply an oscillation of air pressure. It is perceived by humans via a complex series of transformations. First, air pressure oscillation is conducted (and certain frequencies partially amplified) via ear canal to bones of middle ear. They transfer oscillations into vibrations of the basiliar membrane that in turn stimulates inner hair cells. By release of neurotransmitters mechanical energy is converted to electrical signals that are further processed by central neural system in which neurons communicate primarily by firing pulses. Increase of stimulation in inner hair cells translates to increase of firing rate of a neuron. There is however a limit to a neuron's firing rate due to its refractor period.

The crucial point is that from computational point of view organ of Corti performs \emph{Fourier transform}. That is, each section of the basiliar membrane is tuned to a narrow frequency band and neural response is commensurate to energy in that band. It is therefore natural to construct phonemic classifiers based on spectral data.

\section{Nonlinear means}

Our primary motivation is to discriminate between phonemes. It is well known that location of formants is an important characteristic of vowels. Thus would like to find ways other than LPC to quantify ``peakness'' of formants in the spectral envelope. One way to do that is to start with arithmetic means 
\begin{align*}
M(x_1, x_2, \ldots, x_n) &= \dfrac{x_1 + x_2 + \cdots + x_n}{n}.
\intertext{and look for their generalizations. One obvious one is to use weighted linear combinations}
L(x_1, \ldots, x_n) &= a_1 x_1 + \ldots a_n x_n
\end{align*}
which leads to approaches like logistic regression or support vector machines. 

We wish to investigate nonlinear functions. In nonlinear algebras, useful simplifying laws such as  commutativity or distributivity of operations do not need to hold. In order to obtain a reasonably small set of functions in nonlinear algebra we will work with functions derived from \emph{symmetric} operators. These are functions $f$ invariant with respect to argument transpositions
$$
f(x_1,\ldots, x_i, x_{i+1}, \ldots)= f (x_1,\ldots, x_{i+1}, x_{i}, \ldots)
$$
and thus also with respect to any permutations of its arguments. In fact, they are uniquely defined, once one specifies them on the whole cone $x_1 \leq x_2 \cdots \leq x_n$.

Let us recall the classically known generalizations of the arithmetic mean. Quadratic mean is given by formula
\begin{align*}
M_2(x_1, x_2, \ldots, x_n) &= \sqrt{ \dfrac{x_1^2 + x_2^2 + \cdots + x_n^2}{n}},
\intertext{the geometric mean by }
G(x_1, x_2, \ldots, x_n) &= \sqrt[n]{x_1 \cdot x_2  \cdots  x_n},
\intertext{and the harmonic mean by}
H(x_1, x_2, \ldots, x_n) &= \Bigl( \dfrac{x_1^{-1} +  x_2^{-1} + \cdots + x_n^{-1}}{n}\Bigr)^{-1},
\end{align*}
The following ordering between these means holds:
\begin{align}
H(\vecx) \leq G(\vecx) \leq M(\vecx)\leq M_2(\vecx). \label{eqn:gah}
\end{align}

One can define still more general means of which arithmetic, quadratic and harmonic means are special cases. Namely, for $\alpha \not = 0$ and $\vecx > 0$ set
$$
M_\alpha(x_1, \ldots, x_n) = \Bigl( \dfrac{x_1^\alpha + \cdots + x_n^\alpha}{n} \Bigr)^{1/\alpha}
$$
One can show that in fact
$$
\lim_{\alpha \rightarrow 0} M_\alpha(x_1, \ldots, x_n) = G(x_1, \ldots,x_n)
$$
so that the geometric mean is also a member of the family.
\begin{proof}
By L'H\^ospital rule we have 
\begin{align*}
\lim_{\alpha\rightarrow 0} \log M_\alpha(\vecx) &= \lim_{\alpha\rightarrow 0}\dfrac{\log \sum_i x_i^\alpha}{\alpha} \\
&=\lim_{\alpha\rightarrow 0} \dfrac{\sum_i x_i^\alpha \ln x_i}{\sum_i x_i^\alpha}\\
&= \dfrac{\sum_i \ln x_i}{n} = \ln \sqrt[n]{x_1\cdots x_n}
\end{align*}
\end{proof}

Inequalities \eqref{eqn:gah} are just special cases of the following theorem:
\begin{theorem}
If $\alpha < \beta$ then for nonnegative $x_1, \ldots, x_n$ we have
 $$M_\alpha(x_1, \ldots, x_n) \leq M_\beta (x_1, \ldots, x_n).$$
\end{theorem}

For $\alpha \not = 1$ the means $M_\alpha$ are not linear, meaning the equality
$$
M_\alpha(\vecx + \vecx') = M_{\alpha}(\vecx) + M_\alpha(\vecx')
$$
does not need to hold. However they are multiplicatively homogeneous, meaning 
$$
M_\alpha ( c\vecx)= c M_\alpha(\vecx).
$$
Since sound data is usually provided in logarithmic scale one may desire to possess means that are additively homogeneous so that 
$$
F (c \cdot (1, \ldots,1) + \vecx)= c + F(\vecx)
$$
A family of such means is given by the formula
\begin{align}
A_\alpha(x_1, \ldots, x_n) &= 
\ln(M_\alpha(\exp(x_1), \ldots, \exp(x_n))) 
\intertext{where in particular}
A_0(x_1, \ldots, x_n ) &= \dfrac{x_1 + \ldots + x_n}n, \label{A0} \\
\lim_{\alpha\rightarrow -\infty}A_\alpha(x_1,\ldots,x_n) &= \min(x_1, \ldots, x_n), \label{Amin} \\
\lim_{\alpha\rightarrow \infty}A_\alpha(x_1,\ldots,x_n) &= \max(x_1, \ldots, x_n). \label{Amax}
\end{align}

\section{Quantiles}

Minimum and maximum arise naturally as the limiting cases of nonlinear means studied in the previous section, since
\begin{align*}
\lim_{\alpha\rightarrow -\infty}  M_\alpha(x_1, \ldots, x_n) = \min(x_1, \ldots, x_n)\\
\lim_{\alpha\rightarrow\infty}  M_\alpha(x_1, \ldots, x_n) = \max(x_1, \ldots, x_n)
\end{align*}
A simple, but important observation is that both $\min$ and $\max$ do not generate new values. In particular evaluation of any expression of variables $x_1, \ldots, x_n$ using only composition of $\min$ and $\max$ operators results in one of $x_1, \ldots, x_n$. A natural class of symmetric operators in the algebra generated by $\min$ and $\max$ are \emph{quantiles}. In fact, under continuity requirements they are the only symmetric $n$-ary operators in that algebra. 

From computer science point of view it is natural to ask how to compute quantiles using a given class of $\min$ and $\max$ gates. In this context one may ask for instance how many gates one needs, or what is the depth of the resulting feed-forward circuit. The answer differ depending on the kinds of $\min$ and $\max$ operators one allows. 

Suppose for instance that one allows $\min$ and $\max$ gates of arbitrary arity. Obviously $n$-ary $\min$ and $n$-ary $\max$ are both representable by one gate. Any other quantile can be represented by circuits of depth two, for instance the second largest element function can be found by computing either of the following two expressions:
\begin{align*}
\min\bigl(\max(x_2, \ldots, x_n), \max(x_1, x_3, \ldots,x_n), \max(x_1,x_2, x_4, \ldots), \ldots\bigr)\\
\max\bigl(\min(x_1,x_2), \min(x_1,x_3), \ldots, \min(x_i, x_j), \ldots\bigr)
\end{align*}

The restriction to using binary $\min$ and $\max$ gates is more delicate. Results about computations of quantiles can be deduced from classical studies on sorting networks. Well-known is bitonic sorting network of Batcher, which yields explicit circuits for quantiles of $n$ variables of depth $O(\log^2n)$. 
Theoretical improvement came in work \cite{AKS} which showed that  sorting networks with $O(n \log n)$ comparators and depth $O(\log n)$ exist, which is asymptotically optimal. However the implied constant is quite large and research in this area is ongoing. 

\begin{figure}[!ht]
\begin{center}
\includegraphics[width=0.7\textwidth]{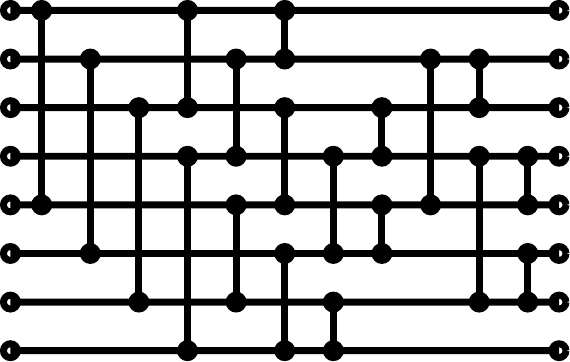}
\end{center}
\caption{Sorting network with 8 inputs and depth 6. Inputs are supplied on the left, outputs (sorted values) can be read out on the right. Each vertical wire signifies reordering of inputs on corresponding horizontal wires.}
\label{fig:sorting8}
\end{figure}

For values of $n$ up to 8, optimal networks are known. For instance, a sorting network with 8 inputs and optimal depth 6 is shown in Figure \ref{fig:sorting8}.

\section{Fuzzy logic}

Mathematical functions introduced so far ($M_\alpha$, $A_\alpha$ and quantiles) are essentially monotonous. When spectral data increase they increase as well. In order to gain \emph{contrast} it is useful to have a nonsymmetric operation of say two variables, ``increasing'' in the first and ``decreasing'' in the second. A well known framework that uses such functions is fuzzy logic. 

Fuzzy logic dates back to works of Lukasiewicz and Post. H. Weyl in 1940 proposed a fuzzy logic where propositions are assigned values in the unit interval. He generalized ordinary Boolean logic operators as follows:
\begin{gather*}
a\text{~and~}b = \min(a,b) \\
a\text{~or~}b = \max(a,b) \\
\text{not~}a = 1 -a \\
a\text{~implies~}b = 1 -a + min(a,b) \\
= min(1, 1-a+b),
\end{gather*}
where $a$ and $b$ take values in the interval $[0,1]$. Many other operators in fuzzy logic have been since proposed. For instance instead one can use $t$-norms or $t$-conorms. Fuzzy logic gained prominence in the foundational paper on fuzzy sets written by L. Zadeh \cite{zad65}. In this paper he used fuzzy logic as a means to represent vague linguistical and cognitive notions.


\section{Eliminating negation}

We note that using min and max in fuzzy logic has been shown to be the only choice  under natural assumptions \cite{MR0414314}. Using negation is however problematic for several reasons.

The first reason arises from implementation issues one encounters when translating fuzzy logic expressions into electronic circuits.  Both negation and implication such as Lukasiewicz implication need to be implemented with active electronic circuits. 

\begin{figure}[!ht]
\begin{center}
\includegraphics[width=0.7\textwidth]{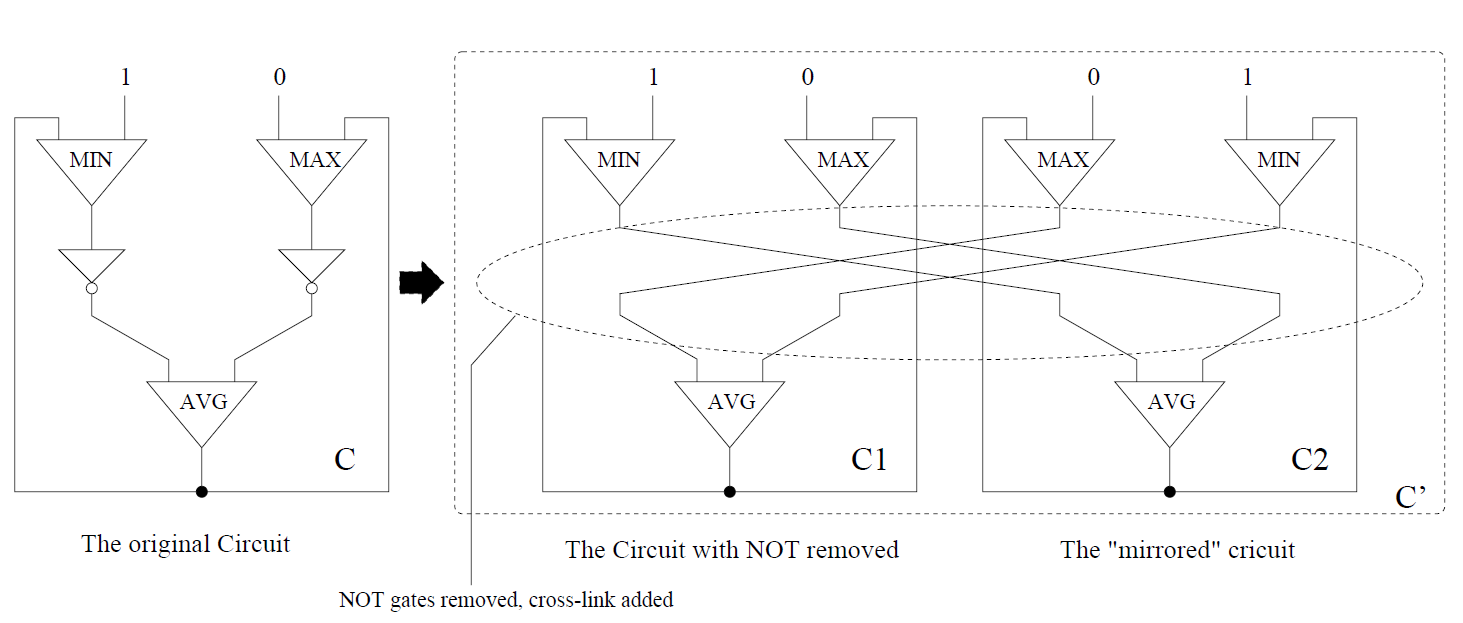}
\end{center}
\caption{How to convert MIN-MAX-NEGATION-AVG circuit to MIN-MAX-AVG circuit}
\label{fig:BlumNegation}
\end{figure}

This issue can be partially addressed. Passive circuits whose inputs are normalized to $[0,1]$ can compute with negation $\bar a = \text{not~}a = 1 -a$, if those are prepared beforehand due to identities
\begin{align*}
\overline{\min(x,y)} &= \max(\overline x, \overline y)\\
\overline{\max(x,y)} &= \min(\overline x, \overline y)
\end{align*} 
This idea is a special case of negation conversion for MIN-MAX-AVG circuits described in \cite{blum} (see Figure \ref{fig:BlumNegation}).

The second reason lies in conflict with our understanding of auditory perception. Introducing negation implies providing lower \emph{and} upper bounds on spectral data. Establishing a lower bound may be reasonable, because it should correspond to the smallest intensity of audible sound. However, an upper bound should probably reflect the largest sound a human ear can sustain without damage, or possibly the intensity threshold at which a harmonic causes the maximum possible firing rate in the auditory nerve. Both of these quantities are very hard to measure and are probably quite variable. It is therefore  unclear why one should let such an upper bound affect information processing in phoneme classification circuits.

A final disagreement is of philosophical nature. Fuzzy logic traditionally deals with uncertain notions primarily of linguistic nature, such as ``tall'', or ``warm''. Spectral content of sound is of course also uncertain, due to uncertainty principle of Fourier transform. However the uncertainty is of a different kind, taking form of a random variable \cite[Chapter 10]{Brockwell}.

%

We thus propose to leave the realm of fuzzy logic and instead use the difference operator (this has value $x-y$ for a pair of real valued variables $x$ and $y$). If this operator is applied to two additively homogeneous expressions such as quantiles or generalized means $A_\alpha$ described earlier, it turns volume sensitive expressions into volume invariant expressions. This invariance is likely to be a desirable feature of classificators. 

We thus arrive at the definition of \emph{KS-algebra} $\MM$. An element of the algebra is any expression of spectral components of sound and the zero value, using operators
\begin{itemize}
\item the minimum $\min(x_1, \ldots, x_n)$,
\item the maximum $\max(x_1, \ldots, x_n)$,
\item the difference $x_1 - x_2$,
\item the additively homogeneous means $A_\alpha$.
\end{itemize}
In view of from \eqref{A0}, \eqref{Amin}, \eqref{Amax}, the addition of operators in the last category  can be seen as constructing a link between linear circuits and $\min$--$\max$ circuits, allowing for direct comparison of performance of linear and nonlinear circuits.

\section{Binary Classifiers arising from  $\MM$}

An interesting experiment is described in work \cite{Hermansky94}, \cite{Hermansky11}. A sentence was processed by a filter with frequency response approximating inverse to spectral envelope of a vowel. In spite of this, the vowel was recognized by speakers. It is thus not the absolute spectral content that determines phonemes, but rather relative. This is confirmed for instance also in work \cite{Prendergast12}.

In our context, a binary classifier will consist of two ingredients. The first one is a function $f$ of KS-algebra algebra $\MM$. The second ingredient describes what one concludes given an evaluation of $f$ on spectral data $\vecs$. We can distinguish the following natural variants of classifiers.


\begin{itemize}
\item A {\bf ${\bf Z}$-classifier}  is described by a function $f\in \MM$. It decides that a phoneme is of the first class if $f(\vecs) < 0$, and decides that is of the second class if $f(\vecs) > 0$.
\item A {\bf${\bf B}$-classifier} consists of  a pair $(f, c)$, where $f \in \MM$ and $c$ is a real number. It decides that a phoneme is of the first class if $f(\vecs) < c$, and concludes that is of the second class if $f(\vecs) > c$.
\item an {\bf ${\bf A}$-classifier} consists of a pair $(f,  c)$, where $f \in \MM$ and $c$ is a real number. It decides that a phoneme is of the first class if $f_1(\vecs) < c$, but makes no conclusion, if this condition is not met.
\end{itemize}
From $A$-classifiers we will distinguish the subclass of $A^+$ classifiers that are charactized by the condition $c\leq 0$.

Let us remark that unlike models with continuous parameters, the classificators are really described by their structure and support, the latter being defined as follows.
\begin{align*}
\supp(0) &= \{\} \\
\supp(s_i) &= \{i\} \\
\supp(\min(f_1, f_2)) &= \supp(f_1) \cup \supp(f_2) \\
\supp(\max(f_1, f_2)) &= \supp(f_1) \cup \supp(f_2) \\
\supp(A_\alpha(f_1, \ldots, f_n)) &= \bigcup_i \supp(f_i)
\end{align*}

\bibliographystyle{elsarticle-num}
\bibliography{alg_bib}

\begin{thebibliography}{10}
\bibitem{Yamakawa93}
T.~Yamakawa, A fuzzy inference engine in nonlinear analog mode and its
  application to a fuzzy logic control, IEEE Transactions on neural networks 4.
\end{thebibliography}

\end{document}